\documentclass[twocolumn]{jpsj2}
\usepackage{epsfig}
\def\gsim{\; $\raise0.3ex\hbox{$>$}\llap{\lower0.8ex\hbox{$\sim$}}$\;}
\def\lsim{\; $\raise0.3ex\hbox{$<$}\llap{\lower0.8ex\hbox{$\sim$}}$\;}

\title{Fractional $S^z$ excitation and its boundstate for  the S=1/2 antiferromagnetic zigzag spin chain in a magnetic field}
\author{Kouichi Okunishi and  Takashi Tonegawa${}^1$}
\inst{
Department of Physics, Faculty of Science, Niigata University, Igarashi 2, 950-2181, Japan\\
${}^1$Department of Mechanical Engineering, Fukui University of Technology, 6-1 Gakuen 3, Fukui 910-8505, Japan.
}
\date{\today}

\abst{ We demonstrate that domain-wall type excitations carrying the fractional magnetization $S^z=\pm1/3$ and their bound state  describe the low-energy physics around the 1/3 plateau of the antiferromagnetic zigzag spin chain in the strongly frustrated region.
In particular we discuss  the relevance of such domain-wall excitations to the characteristic properties of the magnetization curves: the 1/3 plateau  accompanying the spontaneous symmetry breaking of the translation, the cusp singularities above and/or below the plateau, and the even-odd effect in the magnetization curve.}

\kword{frustration, 1/3 plateau, cusp, even-odd effect, domain wall, fractional $S^z$}

\begin{document}
\maketitle

\section{introduction}

In order to understand the role of the frustration in low-dimensional quantum spin systems, the $S=1/2$ antiferromagnetic zigzag spin chain has been one of the most fundamental quantum spin models.\cite{mg,zigzagss,halzig,tone1,igarashi,nomura,allen,sen,whiteaffleck} The Hamiltonian of the model is given by
\begin{eqnarray}
{\cal H}= \sum_{i}[J_1 \vec{S}_{i}\cdot\vec{S}_{i+1} +J_2
\vec{S}_{i}\cdot\vec{S}_{i+2} ] - H \sum_i S_i^z,
\label{zigzag}
\end{eqnarray}
where $\vec{S}$ is the $S=1/2$ spin operator, $H$ is the magnetic field, and $J_1$ and $J_2$ denote the nearest and next nearest neighbor couplings respectively. 
We also introduce a notation $\alpha=J_2/J_1$ for simplicity.  
A remarkable point of the Hamiltonian (\ref{zigzag}) is that, although its form is very simple,  it captures some characteristic aspects of physics induced by the frustration; 
At zero magnetic field, the well-known transition between the spin fluid and dimer phases occurs at $\alpha\simeq 0.2411$\cite{nomura}. Moreover, the ground state is exactly represented as the dimer state at $\alpha=0.5$, which is called Majumdar-Ghosh point.\cite{mg}
In addition, we should note that the zigzag chain is realized as SrCuO$_2$\cite{matsuda}, Cu(ampy)Br$_2$\cite{kikuchi}, (N$_2$H$_5$)CuCl$_3$\cite{haginaru} and F$_2$PIMNH\cite{hosokoshi}.

\begin{figure}[ht]
\begin{center}
\epsfig{file=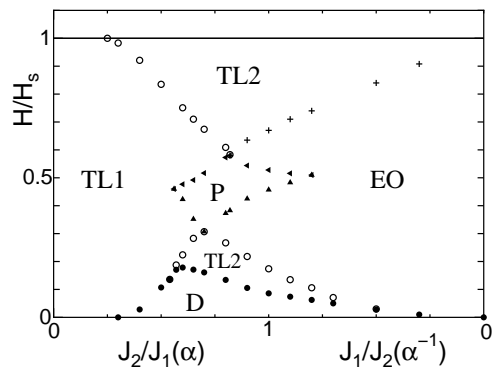,width=6.5cm}
\end{center}
\caption{Magnetic phase diagram of the zigzag spin chain.\cite{OT1}  D: dimer
gapped phase, P:1/3 plateau, TL1: one component TL liquid, TL2: two
component TL liquid, and EO: even-odd behavior blanch.  The open
circles denote the position of the cusp singularities. The solid
circles are the boundary of the dimer gapped phase. The triangles mean
the upper and lower edge of the 1/3 plateau, and the crosses indicate
the upper edge of the even-odd effect in the magnetization curve. The
saturation field is normalized to be unity.} \label{phase}
\end{figure}

Recent evolution of the zigzag chain physics is that the magnetic phase diagram is actually obtained in Ref.[\cite{OT1}].
Of course, a lot of efforts had been made so far,\cite{tone2,schmidt,cabra,cusp,maeshima} but it was a  difficult problem to illustrate reliable magnetization curves in the strongly frustrated region.
As is shown in FIG.\ref{phase}(see also FIG.\ref{fmh1}), the obtained  phase diagram reveals that a variety of exotic behaviors is cooperatively induced by the frustration and the magnetic field.
For $0.56\lsim \alpha\lsim 1.25$\cite{pedge}, the magnetization plateau appears at 1/3 of the full moment, accompanying the spontaneous breaking of the translational symmetry by period three.
Moreover, the cusp singularities in the magnetization curve show quite interesting behaviors around the 1/3 plateau; As $\alpha$ is increased, the high-field cusp merges into the 1/3 plateau at $\alpha \simeq 0.82$. 
Also the low-field cusp merges into  the 1/3 plateau at $\alpha \simeq 0.7$, but it appears again when $\alpha > 0.7 $.
Here we note that the behaviors of the cusps near the saturation field or the lower critical field  are explained well by the shape change of the spin wave or spinon dispersion curves, from which  the two component Tomonaga-Luttinger(TL) liquid can be expected for TL2 in FIG.\ref{phase}.\cite{cusp,mobius}
In addition, we find an interesting even-odd effect in the magnetization curve of a finite size system for $\alpha > 0.7 $.

A series of the above intrinsic structural changes of the magnetization curves indicates that the zigzag chain still contains quite rich physics.
In order to discuss the microscopic mechanism behind such exotic magnetic phase diagram, we would like to emphasize the following features of the zigzag chain:
\begin{itemize}
\item  Frustrating interaction: Usually, the frustration effect is discussed in the context of the triangular structure of the lattice. 
However, here we want to stress the competition between the single chain structure($J_2\gg J_1$) and double chain structure($J_2 \ll J_1$) of the zigzag lattice, which provides some essential insights for the frustration effect in the zigzag chain.

\item Translational invariance:  the Hamiltonian (\ref{zigzag}) connects the above two limits continuously without loss of the translational invariance. 
The parameter $\alpha$ interpolates the single and double chain natures continuously.

\item The 1/3 plateau is extending up to the Ising anisotropic limit, where the $\uparrow\uparrow\downarrow$ order is realized. This means that the $\uparrow\uparrow\downarrow$ order in the Ising limit can be a good starting point to discuss  the isotropic case. 
\end{itemize}

On the basis of the above noted features of the zigzag chain, in the following, we want to review that the domain-wall(DW) excitations carrying fractional value of $S^z$ successfully explain the characteristic behaviors of the magnetization curves.\cite{OT2} 
In \S 2 and \S 3 we briefly overview the important features of the magnetization curves and  the 1/3 plateau state for the zigzag chain of some typical parameters\cite{OT1}.
In \S 4, we introduce the zigzag XXZ chain, which captures more clearly the above-noted three key points, and then explain the essence of the fractional-$S^z$ DW excitations. 
In \S 5, we describe the implications of the DW picture for the magnetization curves and finally summarize the conclusions  in \S 6

\section{ magnetization curves}

\begin{figure}[htb]
\begin{center}
\epsfig{file=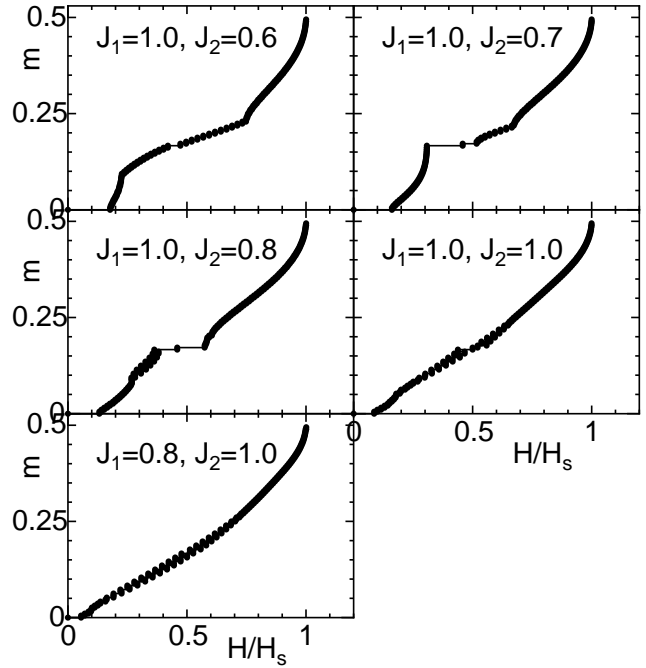,width=8.5cm}
\caption{Magnetization curves for $\alpha=0.6$, 0.7, 0.8, 1.0 and $\alpha^{-1}=0.8$.}\label{fmh1}
\end{center}
\end{figure}

In FIG.\ref{fmh1}, we show the magnetization curves for $\alpha=0.6, 0.7, 0.8,1.0$ and $\alpha^{-1}=0.8$, which were obtained with the DMRG\cite{dmrg} for 192 spins.
In the figures, the scale of the magnetic field is normalized by the
saturation field $H_s$ and $m$ is the magnetization per spins.
We also define $M$ as the total $S^z$ of the system for later convenience. 
First of all, we can see that the 1/3 plateau  appears for $0.56\lsim \alpha\lsim 1.25$,  accompanying the spontaneous symmetry breaking of the translational invariance.  
Here, we note that an anomalous step in the 1/3 plateau for $\alpha=0.7$ and 0.8 is due to the open boundary effect.  
In addition,  the cusp singularities can be seen in low and high field regions.
A more interesting point is that these cusps behave interestingly around the 1/3 plateau;
At $\alpha=0.7$,  the low field cusp merges into the 1/3 plateau and the high field cusp also merges into the $1/3$ plateau near $\alpha\simeq 0.8$.  
Moreover, for $\alpha>0.7$, we can find that anomalous even-odd oscillation with respect to the magnetization $M$.  
For $\alpha>0.82$ the even-odd behavior is extended above the 1/3 plateau.
As $\alpha$ is increased further, the shape of the magnetization curve approaches that of the (double)Heisenberg chain.  
In the figure of $\alpha^{-1}=0.8$, the magnetization curve becomes  continuous around $m=1/3$, implying that there is no $1/3$ plateau.  
Finally in $\alpha^{-1}\to 0$ limit, the magnetization curve approaches to that of the Heisenberg chain.
Here we should make a comment on the even-odd oscillation of the magnetization curve. 
The overhanged steps corresponding to $M$=odd must be skipped in the true magnetization curve of the finite size system.  
However we have shown them, since they are reflecting an interesting aspect of the double-chains nature of the system.

The properties of the cusp singularities near the saturation field or the lower critical field  are explained well by the shape change of the spin wave or spinon dispersion curves.\cite{cusp,mobius}
However, these previous picture of the cusps is not satisfactory for a series of the subtle behaviors of the cusps around the 1/3 plateau.

\section{1/3 plateau state}

\begin{figure}[t]
\begin{center}
\epsfig{file=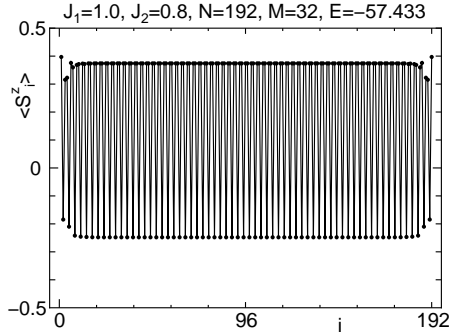,width=6cm}
\end{center}
\caption{ The distribution of the local spin moment at $M$=32(1/3
plateau) for $\alpha=0.8$\cite{OT1}}\label{spindistribution}
\end{figure}

Let us discuss the nature of the 1/3 plateau, which is the starting point of analyzing  the  microscopic origin of the magnetic phase diagram.
According to the quantization condition of the magnetization plateau\cite{oshikawa}, the 1/3 plateau state satisfies $q(1/2-m)=$integer with $m=1/6$, where $q$ is the period of the plateau state.   
This implies that the translational symmetry of the Hamiltonian (\ref{zigzag}) must be spontaneously broken by $q=3,6,\cdots$. 
The spin profile of the plateau along the chain direction is shown in FIG.\ref{spindistribution}, where the ``$\uparrow\uparrow\downarrow$'' type long-range order is realized.  
Moreover, the boundary effect decays rapidly, which supports that the 1/3 plateau state has the excitation gap.

Since the zigzag chain is a system having both of the geometrical frustration and the quantum fluctuation, one may wonder why such a classical $\uparrow\uparrow\downarrow$ order is still standing against the  strong quantum fluctuation and the frustration. 
This may be a very naive but important question to understand  the zigzag chain.
An important point is that the 1/3 plateau phase is extending up to the Ising anisotropic limit, for which the $\uparrow\uparrow\downarrow$ array was proven in Ref.[\cite{morihori}].
The triangular structure of the lattice stabilizes the 1/3 plateau of the $\uparrow\uparrow\downarrow$ order in wide parameter region in the classical limit.\cite{miya}

An qualitative explanation for the stability of the $\uparrow\uparrow\downarrow$ order in the quantum case may be possible by considering the role of the XY term.
Since the XY term tries to change a $\uparrow\downarrow$ pair on a bond into a singlet, the influence of the quantum fluctuation is the most significant at the zero magnetic field(subspace of  $M=0$), where the number of the  $\uparrow\downarrow$ pair is maximum.
At the 1/3 plateau, however,  the number of the $\uparrow\downarrow$ pair is diluted by the  $\uparrow\uparrow\downarrow$ so that the effect of the quantum fluctuation should be  reduced  certainly.
Here we should recall that,  in the $M=0$ subspace of the usual XXZ chain,  the classical N\'eel order survives up to the isotropic limit.
Thus it may be natural that the  $\uparrow\uparrow\downarrow$ order at the 1/3 plateau of the zigzag XXZ chain survives against the quantum effect up to the XY-like region.\cite{tone3}

\section{domain wall excitations}

In order to discuss the nature of the excitations around the 1/3 plateau,we introduce the Ising-like XXZ chain with the nearest neighbor(NN) coupling $J_1$ and the next-nearest neighbor(NNN) one $J_2$ in a magnetic field $H$:
\begin{eqnarray}
{\cal H}=  {\cal H}_0 + {\cal H}_{\rm zeeman},
\label{zigzagxxz}
\end{eqnarray}
where ${\cal H}_0\equiv {\cal H}_{\rm NN}+{\cal H}_{\rm NNN}$ is the Hamiltonian of the system part and ${\cal H}_{\rm zeeman}\equiv  -H \sum_{i=1}^{6N} S^z_i$ is the Zeeman term.
The Hamiltonian of the NN and NNN  interactions are defined as
\begin{eqnarray}
 {\cal H}_{\rm NN}&=& J_1 \sum_{i=1}^{6N} h_{i,i+1} , \quad  {\cal H}_{\rm NNN}=  J_2\sum_{i=1}^{6N} h_{i,i+2}, \\
 h_{i,j} &=& \varepsilon (S^x_iS^x_{j}+S^y_iS^y_{j} )+S^z_iS^z_{j}
\end{eqnarray}
where $\varepsilon$ denotes the anisotropy of the XY term.  We assume the system of $6N$ sites so that $M=N$ corresponds to the 1/3 plateau.
The excitation of the usual XXZ chain at the zero field is often called ``spinon'', which is a domain-wall type excitation carrying $S^z=1/2$.\cite{ishimurashiba,takta}
For the present 1/3 plateau of the zigzag chain, we show that the DW excitations carrying fractional $S^z=\pm 1/3$ and their bound state can successfully explain the intrinsic aspects of the low energy physics.\cite{OT2}

Let us first consider the classification of the excitations on the 1/3 plateau in the Ising limit. 
At the 1/3 plateau, one of three types of $\uparrow\uparrow\downarrow$ orders is selected, according to the spontaneous symmetry breaking of the translation.
Then the excitations on such one-dimensional order are described well by the DWs constructed by  combinations of 3 types of the  $\uparrow\uparrow\downarrow$ arrays. 
In Table 1, we list up 6 possible DWs, which carry fractional value of the magnetization.

\begin{table}[bt]
\begin{center}
\begin{tabular}{crc}
 spin array   &  $S^z$       &energy   \\ \hline
 $\cdots\uparrow\downarrow\uparrow\uparrow\uparrow\downarrow\cdots$ & 1/3   &$ \frac{1}{3}(J_1+J_2)$\\ 
 $\cdots\downarrow\uparrow\uparrow\uparrow\downarrow\uparrow\cdots$ & 1/3   &$  \frac{1}{3}(J_1+J_2)$\\
 $\cdots\downarrow\uparrow\uparrow\uparrow\uparrow\downarrow\cdots$ & 2/3   &$  \frac{2}{3}(J_1+J_2)$\\
 $\cdots\uparrow\downarrow\uparrow\downarrow\uparrow\uparrow\cdots$ & -1/3  &$  \frac{1}{3}(-J_1+2J_2)$\\
 $\cdots\uparrow\uparrow\downarrow\uparrow\downarrow\uparrow\cdots$ & -1/3  &$  \frac{1}{3}(-J_1+2J_2)$\\
 $\cdots\uparrow\uparrow\downarrow\downarrow\uparrow\uparrow\cdots$ & -2/3  &$  \frac{1}{3}(J_1-2J_2)$ \\ \hline
\end{tabular}
\end{center}
\caption{The low energy excitation on the 1/3 plateau state, which is described by the domain wall made from three types of the N\'eel ordered spin aligns. These domain-walls can be regarded as  quasi-particles carrying fractional $S^z$. }
\end{table}

\begin{figure}
\begin{center}
\epsfig{file=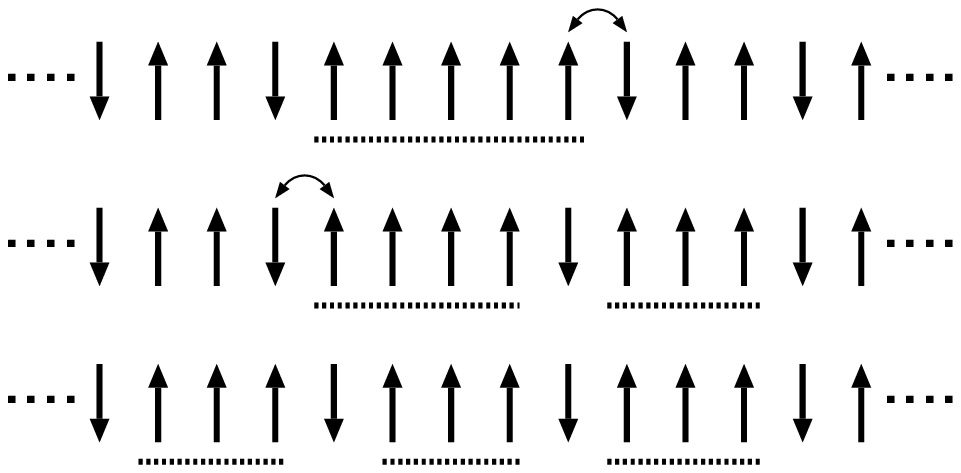, width=5cm}
\end{center}
\caption{ $S^z=1/3$ DW excitation in the Ising limit.}
\label{3dw}
\end{figure}

In order to see the origin of the fractional value of the magnetization, let us start with the single spin flipped state in the 1/3 plateau, which is shown in FIG.\ref{3dw}.
The total magnetization of this state is increased by 1 from the 1/3 plateau.
Using the perturbation of the XY term,  we can systematically construct the low-energy excitations;
by exchanging the nearest neighbor $\uparrow\downarrow$ spin pair, we can see that the five up spin cluster is finally decomposed into three $\uparrow\uparrow\uparrow$ clusters,  which are identical to the $S^z=1/3$ DW in Table I.
 Since these three DWs are equivalent, we can see that each DW carries $S^z=1/3$. 
Moreover, the  $\uparrow\uparrow\uparrow\uparrow$ cluster  in FIG.\ref{3dw} is decomposed into two $\uparrow\uparrow\uparrow$ DWs, implying that the  $\uparrow\uparrow\uparrow\uparrow$ DW in Table I is regarded as  a bound state of two  $S^z=1/3$ DWs.

We  analyze the one-body problem of the $S^z=1/3$ DW in the infinite length chain within the first order of $\varepsilon$. 
Although the higher order terms are required for a quantitative analysis of the problem, we can capture the intrinsic property of the DW excitation within the first order theory.
Let us label the position of the $S^z=1/3$ DW by $x$ which is defined as a site of the center of three up spins.
Then the matrix element of ${\cal H}_0$ in the low-energy region is obtained as
\begin{eqnarray}
{\cal H}_0 | x \rangle =\frac{J_1+J_2}{3} |x\rangle + \frac{\varepsilon J_1 }{2}\left( |x-3 \rangle+|x+3\rangle \right).
\end{eqnarray}
Here an important point is that the NNN term generates only higher energy configurations accompanying the energy rise of order $J_1$ or $J_2$. 
Thus we can obtain the low-energy dispersion curve  of the single DW as
\begin{equation}
\omega(k)= \frac{1}{3}(J_1+J_2)+ \varepsilon J_1\cos( 3 k) .\label{fdwdisp}
\end{equation}
As was seen above,  the NNN terms can not contribute to the low-energy dynamics of the single DW sector within the first order of $\varepsilon$.
Then a natural question arises: what is the role of the NNN term?
In order to see it, we examine the effect of the NNN term on the DW bound state of $S^z=2/3$.
We define the state of the DW bound state at a position $z$ by
\begin{equation}
|z\rangle\equiv \cdots \downarrow\uparrow\uparrow\downarrow\uparrow\!\!\!\>\mathop{\uparrow}_{z}\!\!\!\>\uparrow\uparrow\downarrow\uparrow\uparrow\downarrow \cdots .
\end{equation}
Then the matrix element of ${\cal H}_{\rm NNN}$ for the DW bound state is obtained as
\begin{eqnarray}
{\cal H}_{\rm NNN} | z \rangle =\frac{2}{3}(J_1+J_2) |z\rangle + \frac{\varepsilon J_2 }{2}\left( |z-3 \rangle+|z+3\rangle \right), \label{updwhop}
\end{eqnarray}
implying  that the DW bound state can move by using the NNN term without  extra energy cost. This is a crucial point on the NNN term. 
When $J_1\gg J_2$, the low-energy excitation on the plateau is basically described by the single DW excitation.
On the other hand, when $J_1 \ll J_2$, we can expect that the role of the DW bound state becomes essential in the low-energy excitation.
The switching mechanism between the single DW excitation and the DW bound state is  clearly associated with the crossover between the single chain nature and double chain nature of the zigzag chain. 

\begin{figure}
\begin{center}
\epsfig{file=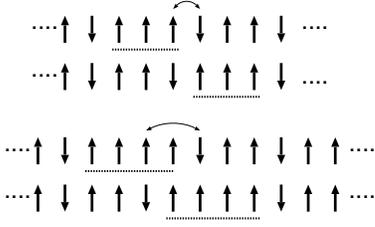, width=5cm}
\end{center}
\caption{Low energy dynamics of the DW excitations.
The NN term($J_1$) shifts the $S^z=1/3$ DW smoothly, while the NNN($J_2$) term move the DW bound state.}
\end{figure}

\begin{figure}[bt]
\begin{center}
\epsfig{file=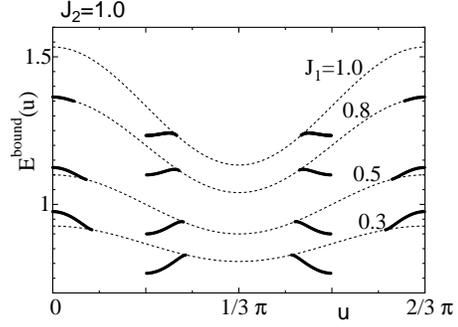, width=6cm}
\end{center}
\caption{The bound state dispersion of $S^z=1/3$ DWs for $\varepsilon=0.1$ and $J_2=1$.\cite{OT2}
The curves are shown for $0<u<2\pi/3$. The solid lines indicate the curves for $J_1=$1.0, 0.8, 0.5 and 0.4.
The dashed lines are the lower bounds of the free two DWs.
} \label{figupdisp}
\end{figure}

We further analyze the two DWs and their bound state problem systematically.
By solving the two DW problem in $S^z=2/3$ sector in the bulk limit, 
we obtain the dispersion curve of the bound state particle
\begin{equation}
E^{\rm bound}(u)=\frac{2}{3}(J_1+J_2)+ \varepsilon J_1 (e^{-3v}+ e^{3v})\cos 3u, \label{bsdisp}
\end{equation}
where $u$ is the momentum of the mass center and $v$ is given by
$
e^{3v}=\alpha \frac{\cos 6u}{\cos 3u}. \label{bsfs}
$
Since the physical solution must satisfy  $e^{3v}\ge 1$ and $-1 \le \cos 3u\le 1$,  the bound state dispersion curve emerges in restricted ranges of $u$.
In FIG.\ref{figupdisp} we  illustrate the  numerically obtained dispersion curve (\ref{bsdisp}) for $\varepsilon=0.1$.
As $J_1$ is decreased, the bound state band comes down to the low energy region.
In particular the bottom of the bound state band becomes lower than that of the free DW band for $\alpha^{-1}\le 0.5$. As will be discussed in the next section, this is important for the switching  of the cusp and even-odd effect of the magnetization curve.

\begin{figure}[bt]
\begin{center}
\epsfig{file=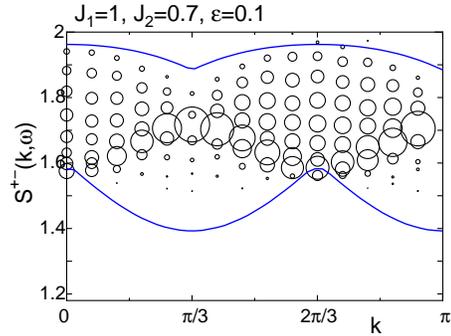, width=6cm}
\end{center}
\caption{Dynamical structure factor $S^{+-}(k,\omega)$ at the 1/3 plateau for $J_1=1.0$, $J_2=0.7$ and $\varepsilon=0.1$. The radius of a circle is proportional to the intensity of the structure factor. Solid lines mean the upper and lower bounds of the three DW continuum. }
\label{figskw}
\end{figure}

In order to see another evidence of the $S^z=1/3$ DW, we have calculated the dynamical structure factor $S^{+-}(k,\omega)$  for $\alpha=0.7$ and  $\varepsilon=0.1 $ at the 1/3 plateau with the exact diagonalization for 30 sites.
Since the $S^z=+1$ excitation is represented as a combination of three $S^z=1/3$ DWs, we can expect the three DW continuum for $\alpha=0.7$.
Figure \ref{figskw} shows the comparison of the  $S^{+-}(k,\omega)$ to the three DW continuum obtained by the 2nd order perturbation of $\varepsilon$. 
Then we can see a good agreement, although the intensity near the lower bound is weak due to the finite size effect.\cite{fsize}

We can also obtain the  bound state dispersion curve of the $S^z=-1/3$ DWs below the 1/3 plateau as
\begin{equation}
E^{\rm bound}(u)=\frac{2}{3}(-J_1+2J_2)+ \varepsilon J_1(e^{-3v}+ e^{3v})\cos 3u,\label{dwnbsdisp}
\end{equation}
where
$
e^{3v}= \frac{ \alpha  \cos 6u -\bar{\Delta}}{\cos 3u}\label{dwnbsev}
$
with $\bar{\Delta}\equiv \frac{-1+2\alpha}{\varepsilon}$ originating form the ``binding energy'' in the zeroth order(Ising limit).  Here we just note that the  $\bar{\Delta}$ term  yields a sensitive $\alpha$ dependence of the bound state dispersion around $\alpha=1/2$\cite{OT2}.

\section{DW excitation and the magnetization curve}

We discuss the relevance of the DW excitations to the magnetization curve.
The magnetization curve of the 1D quantum spin system is generally described  by the hard-core bosonic particle picture;\cite{sorensen,oha} a particle having a  magnetization fills its dispersion curve up to the ``chemical potential'' or ``fermi level'' corresponding to the external magnetic field.
Then the magnetization curve is interpreted as the chemical potential v.s. particle-number curve, where the shape of the dispersion curve is essentially important to figure out the feature of the magnetization curve.
As was seen in the previous section, the excitations around the 1/3 plateau are described well by the DW particles carrying $S^z= \pm 1/3$ and their bound states having $S^z=\pm 2/3$.
Since the total magnetization of the system  always takes  an integer value particularly for a finite size system, the magnetic excitations of $S^z=\pm 1$ around the plateau are described by  combinations of the fractional-$S^z$ DW excitations.

We first discuss the relation between the $S^z=+1/3$ excitation and the characteristic properties of the high-field branch(above the 1/3 plateau).
If $J_1\gg J_2$, the effect of the NNN term is not so big.
Thus the low-energy excitation around the plateau state is basically described by the free $S^z=1/3$ DW excitation.
The magnetization curve near the 1/3 plateau is reflecting the shape of $\omega(k)$ for the  $S^z=1/3$ DW;
the magnetization raises from the 1/3 plateau with the square-root behavior associated with the curvature around the bottom of $\omega(k)$.

\begin{figure}[t]
\epsfig{file=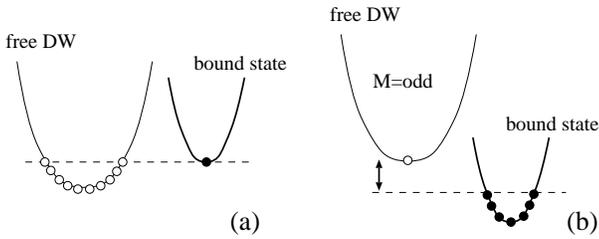, width=8cm}
\caption{Schematic diagram (a) for the cusp singularity and (b) for the even-odd oscillation of the magnetization curve. The open circle denotes the DW excitation and the solid circle denotes the DW bound state. The broken lines indicate the ``chemical potential(fermi level)'' corresponding to the magnetic field}
\label{schem1}
\end{figure}

As $J_2$ is increased and the NNN effect becomes more significant,  the dispersion curve of the DW bound state comes down to the low-energy region.
Then there are two possible situations: the bottom of the DW-bound-state dispersion curve is slightly higher than that of the single DW dispersion curve(FIG.\ref{schem1}-(a)), and the opposite (FIG.\ref{schem1}-(b)).
Here it should be remarked that the switching point between the situations in FIG.\ref{schem1} (a) and (b) for the isotropic case  corresponds to $\alpha\simeq 0.82$, at which the cusp lines meet the higher field edge of the 1/3 plateau in FIG.\ref{phase}.

For the case of FIG.\ref{schem1}-(a) , the magnetization increases along the single DW dispersion curve, as long as the fermi level is below the bottom of the bound state dispersion curve. 
However, when the fermi level touches the bottom of the bound-state dispersion curve,  the magnetization curve captures the band edge singularity to have the cusp(FIG.\ref{schem1}-(a)).
After the fermi level exceeds the bottom of the bound-state dispersion curve, the magnetization can increases rapidly by using the bound-state dispersion curve.
In the context of number of the crossing points of the fermi level and the dispersion curves, we can see that  the one component TL liquid is realized between the field cusp and the 1/3 plateau, where the fermi level intersects only the single DW dispersion. 
On the other hand,  the two component TL liquid is expected above the cusp singularity, since the fermi level crosses both of the single DW and bound state bands\cite{2tl}.

For the case of FIG.\ref{schem1}-(b), the magnetization curve increases by using the bound state dispersion curve.
Here we recall that the excitation of $S^z=1$ is always represented as a combination of three DW particles and thus there are $3|M-N|$ number of $S^z =1/3$ DWs in the system.
Then, if $M=$odd for $N=$even, $3|M-N|-1$ number of the DWs can conform the bound state of $S^z=2/3$, but the remaining one DW can not find its partner.
Thus the remaining one DW has to sit on the free DW dispersion curve, as in the diagram of FIG.\ref{schem1}-(b).
Therefore the $M=$odd state has a slightly higher energy due to the gap between the bottom of the single DW band and the fermi level lying in the bound state band.
On the other hand, for the case of $M=$even, all of the DWs can find their partners and conform the bound states successfully.
Clearly this is the origin of the even-odd behavior of the magnetization curve, for which the one component TL liquid composed of the DM bound state can be expected, since the fermi level crosses only the bound state band.\cite{oddimp}

The analysis of the low-field branch can be done by almost the same line of the argument.
We can easily see that for  $\alpha \ll 1/2 $ the shape of the bound state dispersion  corresponds to the case (a), while  $\alpha \gg 1/2 $  corresponds to the case (b).
However, in contrast to the high-field branch, the $S^z=-2/3$ DW bound state has the zeroth order binding energy, namely $\bar{\Delta}$ term,  which yields the sensitive behaviors of the dispersion curve of the bound state near $\alpha=1/2$.
We have calculated the magnetization curve of $\varepsilon=0.1$ around $\alpha\simeq 1/2$ intensively, and verified that the magnetization curve actually captures the sensitiveness of the dispersion curve.\cite{OT2}

As seen above, the fractional-$S^z$ DW excitation picture can successfully explain the magnetization curve calculated by the DMRG.
Since the dispersion curves in \S 4 are based on the first order perturbation, the higher order contribution of $\varepsilon$ may affect the shape of the dispersion curves.
In order to check this point, we have also performed DMRG calculations for various $\varepsilon$ and verified that the 1/3 plateau and the cusp singularities in the Ising limit are adiabatically connected with the isotropic case;
the topology of the magnetic phase diagram around the 1/3 plateau is almost the same as the present Ising-like XXZ model.
Thus the DW picture is also maintained against the quantum fluctuation of the XY term.

Further we should  make a comment on a connection to the other pictures of the magnetization curve. 
A good picture to describe the magnetization curve near the saturation field is the spin wave excitation from the fully polarized state, the shape change of which explains the cusp singularity successfully\cite{cusp}.
Similarly, the magnetization curve near the zero magnetic field can be also explained by the shape change of the dispersion curves of the DW in the dimer-singlet states for the isotropic case($\varepsilon=1$)\cite{zigzagss,igarashi,brehmer}, or of the dispersion of the DW in the N\'eel type order for the sufficiently Ising anisotropic case.\cite{igarashi}
These excitations from the saturation field/zero-magnetic field limits and the fractional DW picture based on the 1/3 plateau are complemental to each other for a quantitative analysis of the magnetization curve.
The relation between these pictures from the 1/3 plateau and the saturation/zero-magnetic fields is a remaining important problem.
In addition, we also note that how the system in $J_1\to 0$ approaches the decoupled  chains limit is also an interesting problem, since the bosonization results may be available.

\section{summary}

We have reviewed the magnetic phase diagram of the $S=1/2$ zigzag spin chain including the strongly frustrated region and its underlying  microscopic mechanism.
Starting from the 1/3 plateau state in the Ising limit, we have constructed the $S^z=\pm 1/3$ DWs and their bound states carrying $S^z=\pm 2/3$.
Then we have emphasized the role of the single chain and double chain structures of the zigzag lattice; 
Whether the system has the single or double chain characters can be labeled by an integer, while the parameter $\alpha$ interpolating them is a continuous real number. 
Thus, as $\alpha$ varies  gradually, the single and double chain natures of the system have to change at a critical value of $\alpha$.
We think that the $S^z=\pm 1/3$ DW excitations and their bound state strikingly demonstrate such switching of the the single and double chain natures of the excitations, which explains the characteristic behaviors of the cusp singularities around the 1/3 plateau. 
However, the ``interplay'' of the integer quantum number and the continuous parameter itself should be independent of the microscopic details of the excitations. 
We think that this fact may be a essential view point behind the exotic structure of the phase diagram.

Although the zigzag chain is a quite simple model,  we think that it captures the intrinsic aspects common in a class of frustrated spin chains having the zigzag type structure. 
Recently,  $Z_3$ criticality in related models is investigated\cite{orignac,fendley,ohtsuka}, where the connection to the 1/3 plateau of the zigzag chain may be seen.
The stability of the 1/3 plateau is also an interesting problem; a perturbation of  the trimarized interaction may induce $\cdots \uparrow$-singlet-$\uparrow$-singlet$\cdots$ type order rather than the $\uparrow\uparrow\downarrow$.\cite{hida}
We hope that our results stimulates further theoretical and experimental studies of physics induced by the frustration and the magnetic field.



We would like to thank T. Hikihara for fruitful discussion.
We also thank K. Hida and I. Affleck for valuable comments. 
This work was partially supported by a Grant-in-Aid for Scientific Research on
Priority Areas (B) and a Grant-in-Aid for Scientific Research (C) (No. 16540332)
from the Ministry of Education, Culture, Sports, Science and Technology of Japan.


\begin{thebibliography}{99}
\bibitem{pedge} This edge of the 1/3 plateau is extracted from the magnetization curve computed with the DMRG. 
Since this transition is expected to be of BKT type with $Z_3$ symmtetry breaking, the excitation gap is exponentially small near the transition point. A precise value of the plateau edge for the small $J_2$ side is obtained as $\alpha_c=0.487$ by the level spectroscopy method\cite{LS,nomura3} in ref.\cite{tone3}.
\bibitem{fsize} Here the finite size effect may be significant because three DW problem in 10 units of $\uparrow\uparrow\downarrow$ is treated.
\bibitem{oddimp} The $M$=odd system is not relevant to the bulk magnetization curve. The remaining single DW for $M$=odd case  may be regarded as a mobile impurity in the TL liquid of the bound state particle.
\end{thebibliography}
\end{document}